\documentclass[a4paper, amsfonts, amssymb, amsmath, reprint,showkeys,nofootinbib,superscriptaddress]{revtex4-2}

\usepackage{graphicx}
\usepackage[normalem]{ulem}
\usepackage{textcomp}
\usepackage{gensymb}
\usepackage{amsmath}
\usepackage{hyperref}
\usepackage{array}
\usepackage{float}

\newcolumntype{L}[1]{>{\raggedright\let\newline\\\arraybackslash\hspace{0pt}}m{#1}}
\newcolumntype{C}[1]{>{\centering\let\newline\\\arraybackslash\hspace{0pt}}m{#1}}
\newcolumntype{R}[1]{>{\raggedleft\let\newline\\\arraybackslash\hspace{0pt}}m{#1}}

\begin{document}

\title{A fast tunable 3D-transmon architecture for superconducting qubit-based hybrid devices}

\author{Sourav~Majumder}
\affiliation{Department of Physics, Indian Institute of Science, Bangalore-560012 (India)}

\author{Tanmoy~Bera}
\affiliation{Department of Physics, Indian Institute of Science, Bangalore-560012 (India)}

\author{Ramya~Suresh}
\affiliation{Department of Physics, Indian Institute of Science, Bangalore-560012 (India)}

\author{Vibhor~Singh}
\email{v.singh@iisc.ac.in}
\affiliation{Department of Physics, Indian Institute of Science, Bangalore-560012 (India)}

\date{\today}


\begin{abstract}

Superconducting qubits utilize the strong non-linearity of the
Josephson junctions. Control over the Josephson nonlinearity, 
either by a current bias or by the magnetic flux, can be a 
valuable resource that brings tunability in the hybrid system 
consisting of superconducting qubits.
To enable such a control, here we incorporate a fast-flux line for 
a frequency tunable transmon qubit in 3D cavity architecture. 
We investigate the flux-dependent dynamic range, relaxation from
unconfined states, and the bandwidth of the flux-line. 
Using time-domain measurements, we probe transmon's relaxation from 
higher energy levels after populating the cavity with 
$\approx 2.1\times10^4$ photons. For the device used in the 
experiment, we find a resurgence time corresponding to the 
recovery of coherence to be 4.8~$\mu$s.
We use a fast-flux line to tune the qubit frequency and demonstrate 
the swap of a single excitation between cavity and qubit mode. 
By measuring the deviation in the transferred population from 
the theoretical prediction, we estimate the bandwidth of the flux line to be 
$\approx$~100~MHz, limited by the parasitic effect in the design.
These results suggest that the approach taken here to implement a 
fast-flux line in a 3D cavity could be helpful for the hybrid devices 
based on the superconducting qubit.

\end{abstract}


\maketitle

Josephson circuits are the ideal candidates to realize a wide range of 
quantum technologies.
Low dissipation and the ability to implement tailored Hamiltonians 
in a quantum circuit have led to a wide range of matured platforms, 
such as quantum-noise limited amplifiers \cite{castellanos-beltran_widely_2007,yamamoto_flux-driven_2008,macklin_nearquantum-limited_2015}, 
circuit-QED systems \cite{devoret_implementing_2004,blais_quantum-information_2007}, 
and hybrid devices \cite{clerk_hybrid_2020}.
While a wide variety of interactions can be implemented by designing the static 
nonlinearity using Josephson junctions \cite{buluta_natural_2011, xiang_hybrid_2013},
a class of interaction Hamiltonians requires the application of 
resonant or off-resonant pumps \cite{flurin_generating_2012,abdo_full_2013,leghtas_confining_2015}.
One such application of c-QED platform is towards the hybrid devices, where it
can be used as an auxiliary mode.
Hybrid systems based on the mechanical oscillators \cite{lahaye_nanomechanical_2009, 
	oconnell_quantum_2010, lecocq_resolving_2015, pirkkalainen_hybrid_2013}, 
electron spins \cite{zhu_coherent_2011, kubo_hybrid_2011}, surface acoustic waves 
\cite{gustafsson_propagating_2014, manenti_circuit_2017, bolgar_quantum_2018}, and magnons 
\cite{tabuchi_coherent_2015} have been investigated. 
Recent developments on the hybrid devices, based on electrostatic 
coupling with the nanomechanical oscillator \cite{viennot_phonon-number-sensitive_2018}, 
and with acoustic resonator \cite{chu_creation_2018,arrangoiz-arriola_resolving_2019}, 
operating in the number resolved limit have further raised the interest 
towards the c-QED based hybrid devices \cite{clerk_hybrid_2020}.

In hybrid devices, often the requirement of high pump power for the enhancement of the parametric coupling, renders
the integration of the c-QED system incompatible. 
In a high power regime, for example, the transmon qubit decouples 
from the cavity mode and gets excited to the unconfined-states \cite{lescanne_escape_2019}.
The critical power necessary to operate the transmon within the few energy-level 
subspace can be increased by an inductive shunt but not without compromising 
the underlying non-linearity \cite{verney_structural_2019}.
Another useful feature would be the ability to adjust qubit-cavity coupling from 
dispersive to resonant limits by rapidly tuning the qubit frequency with magnetic 
flux. While such fast-flux bias lines are straightforward to design in planer 
devices, integrating them into the 3D-cavity is challenging
\cite{gargiulo_fast_2021, reshitnyk_3d_2016}.


%
With these challenges in mind, investigating the performance of transmon qubit in 3D 
architecture with a fast-flux line could still have practical importance 
\cite{yuan_large_2015,noguchi_ground_2016,gunupudi_optomechanical_2019, 
	peterson_ultrastrong_2019, ofek_extending_2016, heeres_implementing_2017}.
For example, consider a low 
frequency mechanical oscillator coupled to a microwave cavity. In such a device, the 
re-thermalization time, defined as the time taken to reach the mean phonon occupation of one 
after initialization to the quantum ground state, can be large due to the high quality factor
of the mechanical oscillator. Therefore in a hybrid device, if the relaxation time of the qubit from 
unconfined states remains smaller than the re-thermalization time, both the systems,
qubit and the mechanical oscillator can be initialized to their quantum ground state
without a significant loss to the state fidelity. Thus, such initialization in 
the quantum limit can be used for controlled interaction between the two modes.

Here we incorporate a fast-flux line for a frequency tunable transmon qubit in 3D
cavity architecture. We investigate three aspects of such design which is required
for the hybrid devices. The dynamic range of the system for various flux bias is
probed first. We measure the timescale associated with the 
recovery of coherence in the system after a strong pump as the transmon relaxes 
from highly excited states and pump photon leaves the cavity. 
The initialization to the ground state is probed by performing vacuum-Rabi 
measurements while varying the delay between the pump and the control signals.
Finally, the fast-flux is used to demonstrate the single excitation swap between
the cavity and transmon mode.

\begin{figure*}
	\includegraphics[width = 110 mm]{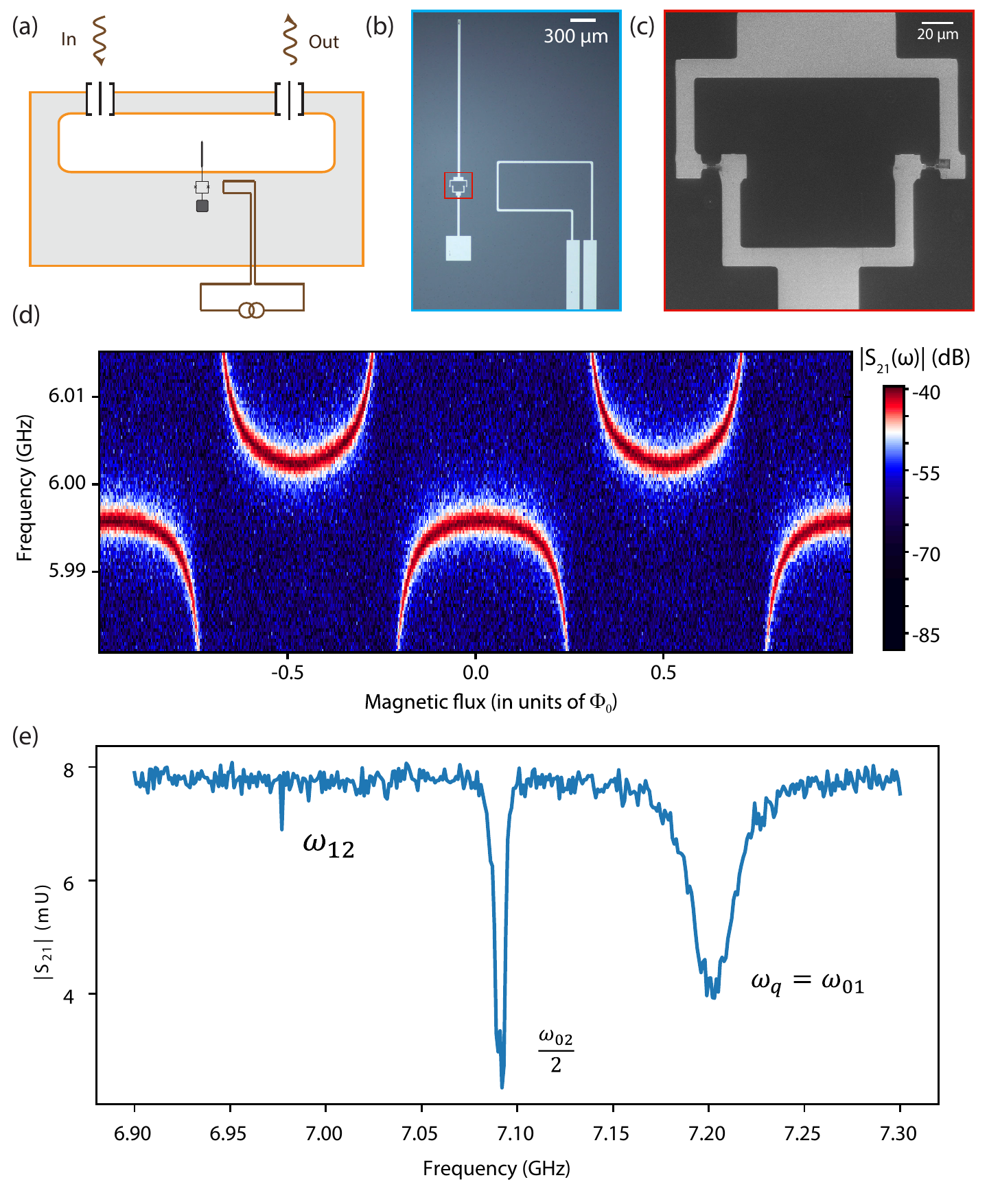}
	\caption{(a) A schematic of the device showing frequency tunable transmon 
		qubit coupled to a 3D cavity. The SQUID loop is positioned in a small 
		recess machined inside the cavity wall to incorporate a port for flux 
		tuning. (b) An optical image of the sample showing the fabricated transmon 
		qubit and the flux line. The thin vertical electrode provides the 
		necessary qubit capacitance and couples to the fundamental mode of the 3D 
		cavity. (c) Scanning electron microscope image of the SQUID loop. 
		(d) Transmission through the cavity $|S_{21}|$ as the magnetic 
		flux threaded by the SQUID loop is varied. (e) Qubit spectroscopy at the 
		high drive power when qubit is flux biased at $\Phi = 0$. Various qubit 
		transitions are labeled accordingly. We determine the qubit anharmonicity to be 
		$\approx$~225 MHz.}
	\label{fig1}
\end{figure*}

Unlike the conventional 3D transmon the position of the SQUID loop is shifted away 
from the center of the cavity into a 
recess created in the cavity wall\cite{paik_observation_2011, juliusson_manipulating_2016}. 
As shown in Fig.~\ref{fig1}(a), the SQUID is shifted to a recess designed inside
the cavity wall. An antenna pointing towards the cavity center provides the 
necessary capacitance to the qubit mode and the coupling to the cavity mode.
The transmon design was simulated using the black-box quantization 
technique \cite{nigg_black-box_2012}. Positioning the SQUID loop in a 
recess allows us to incorporate a local flux line near the SQUID loop 
to tune the qubit frequency rapidly. Such integration is also compatible 
with high coherence cavities designed from superconductors \cite{gargiulo_fast_2021, 
	reshitnyk_3d_2016, navau_long-distance_2014}.

The flux control line is designed to avoid any perturbation to the cavity mode 
while minimizing the relaxation of qubit mode to the flux-drive port. 
Fig.~\ref{fig1}(b) shows an optical microscope image of the fabricated device 
using the shadow evaporation technique on an intrinsic silicon substrate. 
Fig.~\ref{fig1}(c) shows the scanning electron microscope image of the SQUID loop.
The patterned device is placed inside an oxygen-free high thermal conductivity 
(OFHC) copper cavity and cooled down to 20~mK. 
Our measurement setup consists of various absorptive/reflective
filters. 
Detailed schematic of the measurement setup is shown in 
the supplementary information (SI).
We use a 1~GHz low-pass reflective filter on the flux-line and place it close to
the flux drive port. The entire cavity assembly is placed inside the magnetic and
infrared radiation shields.

\begin{figure*}
	\includegraphics[width = 120 mm]{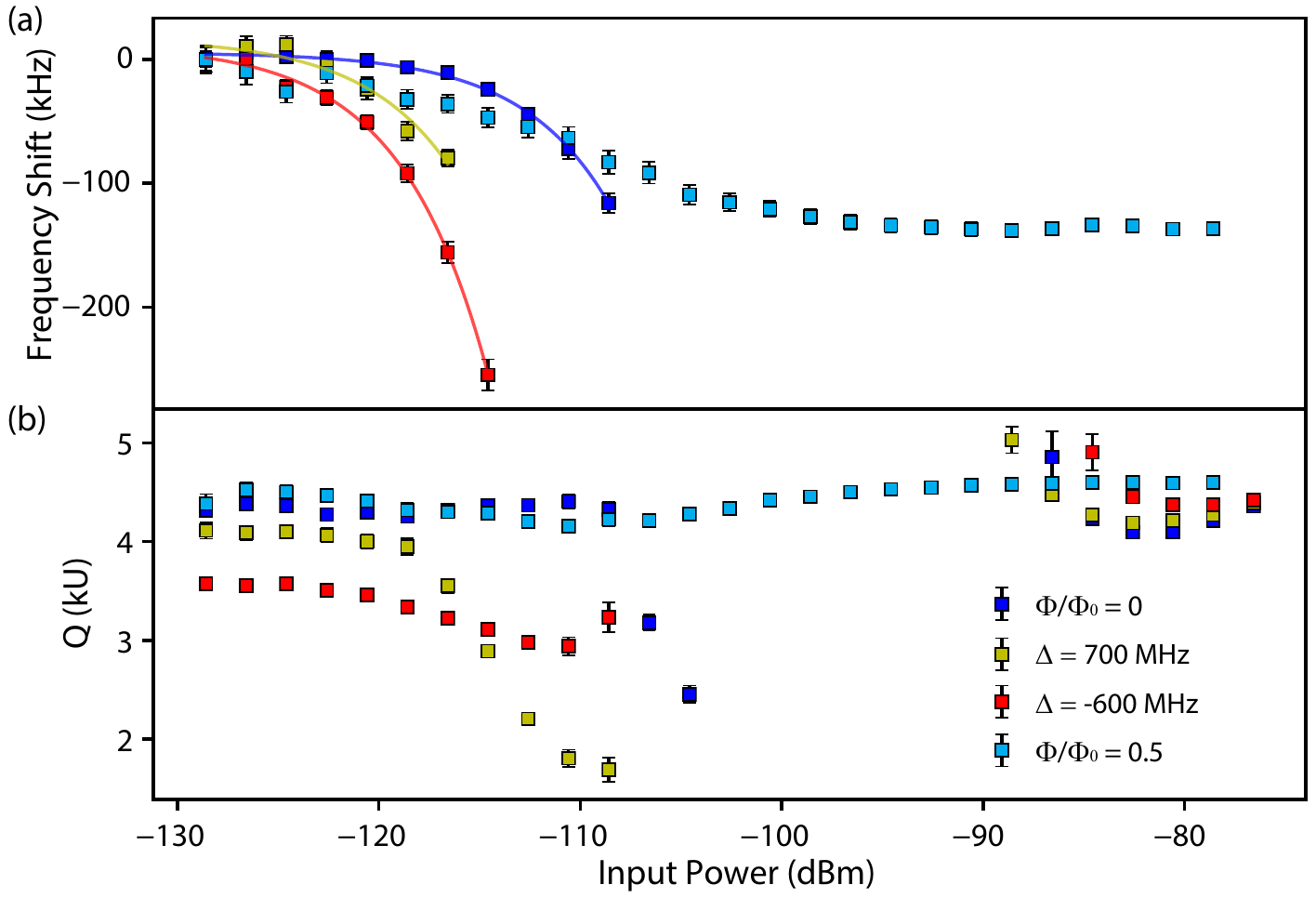}
	\caption{(a) Shift in the dressed cavity frequency as the input power to the 
		cavity is increased for variable magnetic flux. The minimum power 
		-129~dBm  corresponds to the mean cavity occupation of 0.17 photons. The 
		solid lines are numerical fits to extract the cavity nonlinearity. (b) The 
		loaded (total) quality factor of the dressed mode for different input power to the 
		cavity at multiple flux bias points.}
	\label{fig2}
\end{figure*}

We begin by performing spectroscopy measurements on the device. Fig.~\ref{fig1}(d)
shows the cavity spectroscopy measurement as the magnetic flux through the SQUID loop is
changed by varying the current through the flux-line. An avoided crossing,
signifying the strong coupling, between the qubit mode and the cavity mode is 
clearly visible. 
From the qubit spectroscopy, shown in Fig.~\ref{fig1}(e), we determine the maximum qubit frequency
(ground to first excited state transition) to be $\omega_{q}^0/2\pi\sim7.203$~GHz
and corresponding dressed cavity frequency for the ground state as 
$\omega_{c}/2\pi\sim5.996$~GHz. We measure the coupling between two modes to be 
$g/2\pi\sim 87$~MHz, which is close to the designed value 
\cite{nigg_black-box_2012}.
While the maximum qubit frequency depends on the total critical current of the two
junctions, the minimum qubit frequency depends on the asymmetry of the two 
junctions.
From the two-tone spectroscopy measurements, we could tracked the qubit frequency
down to 4~GHz while tuning it with the flux.
Various device parameters are summarized in Table \ref{table}.

At low probe power, the cavity frequency shifts to a 
dressed frequency due to its interaction with the qubit.
Beyond a critical power, the cavity jumps to its bare frequency defining the 
dynamic range of the system. 
In this limit, the phase difference across the junctions evolves continuously.
This has been attributed to the excitation of the qubit to the unconfined 
states lying outside the cosine potential well \cite{lescanne_escape_2019}.
We use scqubits package to compute the energy eigen-states using the 
measured device parameters and find that there are approximately 10 confined
states within the cosine potential well \cite{groszkowski_scqubits_2021}.
The higher transmon levels exhibit larger charge dispersion \cite{koch_charge-insensitive_2007}.
At large probe powers, the higher transmon levels become important,
and the coupled system must be treated by including the Kerr-nonlinearity 
terms. In the dispersive limit, the Hamiltonian of the system can be written as 
$\hat{H}_{sys}/\hbar = \omega_c \hat{a}^{\dagger}\hat{a}+\omega_q 
\hat{b}^{\dagger}\hat{b} - 
\frac{1}{2}\alpha_c\hat{a}^{\dagger}\hat{a}^{\dagger}\hat{a}\hat{a} -
\frac{1}{2}\alpha_q\hat{b}^{\dagger}\hat{b}^{\dagger}\hat{b}\hat{b} +
\chi\hat{a}^{\dagger}\hat{a}\hat{b}^{\dagger}\hat{b}$, where $\omega_c$ 
($\omega_q$) is the cavity 
(transmon) frequency, $\alpha_c$ ($\alpha_q$) is the cavity (transmon) 
Kerr-nonlinearity and $\chi$ is the dispersive shift. 
Due to the qubit-induced Kerr-nonlinearity, even in the dispersive regime the dynamic range of the cavity 
gets limited significantly.
Fig.~\ref{fig2}(a) shows the experimental results of the change in the dressed 
cavity frequency with probe power at the device. 
As the probe power is increased, the dressed cavity frequency changes as
$\omega_c(\bar{n})=\omega_c(0) -2\alpha_c\bar{n}$ \cite{leghtas_confining_2015}, 
where $\bar{n}$ is average number of photons in the cavity.
From an independent calibration of $\bar{n}$ using ac-Stark shift, we deduce 
$\alpha_c$ for different qubit detunings. Additional dataset on ac-Stark shift 
is included in SI.
For zero magnetic flux ($\Phi~=~0$), when qubit detuning 
$\Delta = \omega_{q} - \omega_{r}\approx2\pi\times$1.2~GHz, we estimated the 
the cavity non-linearity to be $\alpha_c/2\pi = -3.2$~kHz. As the qubit mode 
is tuned closer to the cavity $\Delta=-2\pi\times$~600~MHz, 
$\alpha_c/2\pi$ increases to $-27.8$~kHz, which is also indicated by the
reduced dynamic range shown in the Fig.~\ref{fig2}(a).
As expected, the maximum dynamic range is achieved when the qubit is detuned 
furthest to the cavity frequency at $\Phi/\Phi_0=0.5$. 
At this flux-operating point, we also observe a reduction in the cavity
frequency resulting from the asymmetry in the critical currents 
of the SQUID junctions. Similar behavior is observed in the corresponding 
quality factor of the dressed mode, as shown in the Fig.~\ref{fig2}(b).

\begin{figure*}
	\includegraphics[width = 140 mm]{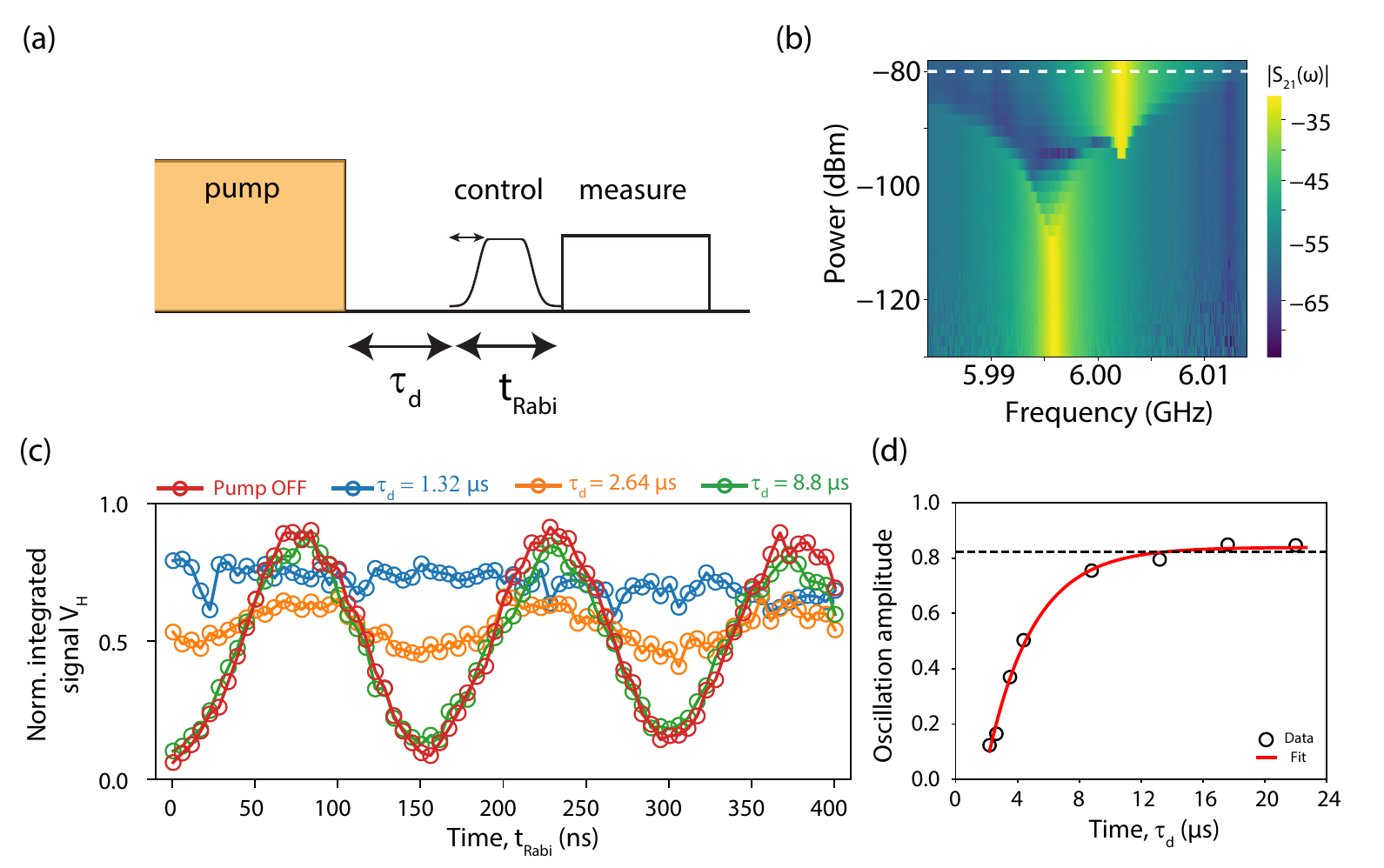}
	\caption{(a) A schematic of the pulse sequence used in the measurement. The 
		cavity is driven with a strong pump pulse, which creates a large 
		population of photons in the cavity and excites the transmon to unconfined 
		states. Subsequently, Rabi measurement protocol with varying delays 
		$\tau_{d}$ is followed. For control, we use a pulse that has a rise and fall 
		parts defined by a gaussian function and central part of the pules is 
		rectangular. The rise (and fall) part of the pulse is defined by a Gaussian
		function of length 35~ns and a $\sigma$ of 9~ns, indicated by
		the arrow.
		(b) Cavity transmission as probe power is increased while biasing the 
		qubit at zero flux quantum. The white dotted line indicates the power used 
		for the high power pulsed measurements. (c) Rabi measurement for different 
		delays between the high power pump and qubit control pulse. 
		A plot of Rabi oscillation without any high power pump is included as a 
		reference. 
		The pump power in the steady-state corresponds to a photon number $n_{d}$ 
		of 2.1$\times10^4$. (d) Amplitude of Rabi oscillation measured for 
		different delays between the pump and the control pulse. 
		Oscillation amplitude when the pump pulse is in off condition is 
		denoted by the black dotted line. Statistical uncertainties from the fits
		are smaller than the marker size.}
	\label{fig3}
\end{figure*}

After the basic characterization of the device, we investigate the high power 
response in the time domain.
The high pump power excites the qubit to unconfined states. It could be 
accompanied by the creation of quasi-particles, which could take a long time to 
relax.
We perform time-domain measurements to probe the resurgence of coherence 
after subjecting the system to a strong pump.

Using the flux-bias, the qubit frequency is tuned to the maximum frequency
 $\Delta=\omega_q-\omega_r\approx2\pi\times$1.2~GHz. 
A control pulse at the qubit frequency is applied. It is followed by a measurement pulse at 
the dressed cavity frequency, corresponding to a steady-state cavity occupation of 
$\approx$~5 photons.
The transmitted signal from the cavity is amplified at 4~K and at room temperature. 
The amplified signal is then down-converted to an intermediate frequency. 
Both quadratures of the IF signal are recorded as a function of time with a 
lock-in amplifier. To improve the readout signal contrast, we average fifty 
thousand time-traces of in-phase and quadrature streams of the readout signal.

Such an ensemble average of time traces can 
then be used to determine the qubit state. We follow an
approach similar to Ref.~\cite{bianchetti_dynamics_2009}
and define a normalized integrated signal 
$V_H$ as $V_H = \frac{1}{2} \frac{\sum_i \left(V_g (t_i) - V_m (t_i)\right)\Delta t }{\sum_i \left(V_g (t_i) - V_s (t_i)\right)\Delta t}$, where $\Delta t$ is the 
resolution of time-axis. $V_g$ ($V_s$) represents the averaged signal 
traces corresponding to the qubit in the ground state (in an 
equal mixture of ground and the first excited state). 
$V_m$ represents signal for the unknown qubit state that is 
being measured. Here, we effectively use the saturation control pulse for 
normalization. It is important to emphasize here that 
$V_H$ slightly deviates from the first excited state probability 
due to loss of population during the measurement process.

\begin{figure*}
	\includegraphics[width = 150 mm]{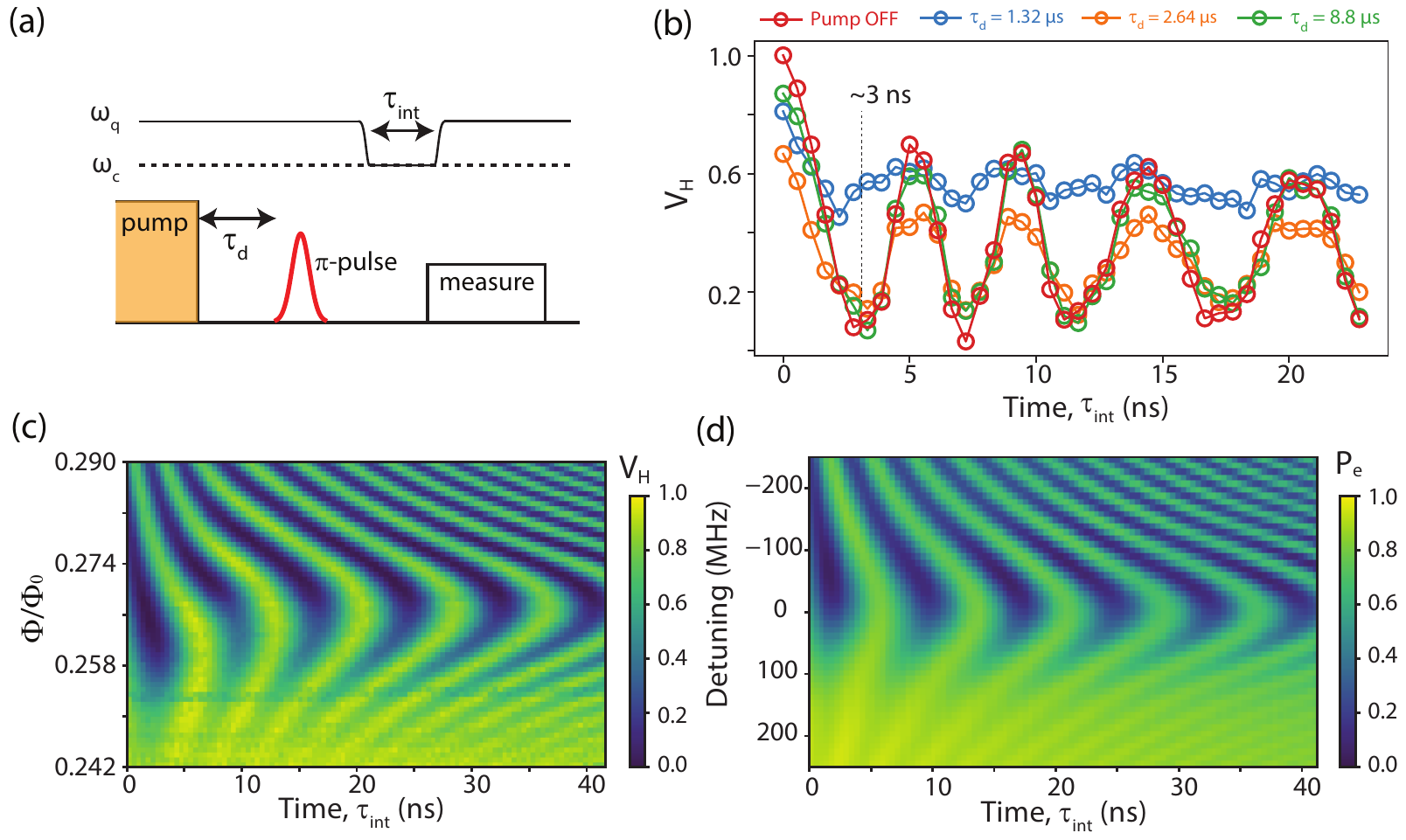}
	\caption{(a) A schematic of the pulse sequence used in the measurement. 
		The interaction  time ($\tau_{int}$) between cavity and qubit is varied for 
		different values of $\tau_{d}$. (b) Normalized integrated signal showing the swap
		of excitation between the qubit and the cavity mode as the interaction length is 
		varied. The black dotted line indicates the time to swap one excitation.
		(c) The colorplot of the normalized integrated signal as the detuning between the 
		cavity mode and qubit is varied. The results were
		obtained from a different device with similar parameters. 
		(d) The colorplot of the simulated qubit population in the excited 
		state while varying the interaction length and relative 
		detuning $\Delta = \omega_q - \omega_c$.}
	\label{fig4}
\end{figure*}

To probe the relaxation of qubit from the higher energy levels to the
ground state, a strong pump tone is applied at the bare cavity frequency, followed by 
the pulsed control and readout scheme as described before. A schematic of the pulse sequence is
shown in Fig.~\ref{fig3}(a).
In the presence of a strong pump, the transmon gets excited to the 
unconfined states resulting in the maximum 
transmission through the cavity at its bare frequency. 
Fig.~\ref{fig3}(b) shows the measurement of transmission 
through the cavity as power of the probe signal is increased. 
The dressed mode shows the characteristic frequency shift due to the presence of 
the qubit. Beyond a critical power, the maximum transmission 
jumps to the bare cavity frequency.
For the pump pulse, corresponding mean occupation of $n_d\sim$2.1$\times$10$^4$
photons in the cavity indicated by the dotted line in Fig.~\ref{fig3}(b).
The calibration of the pump photons is performed by using the ac Stark measurements made at
low probe powers. We have calibrated the total attenuation in the line and calculated the 
total number of photon with respect to bare cavity frequency.
This strong pump pulse excites the qubit to higher unconfined states. By varying the
length of the qubit control pulse, we perform the vacuum Rabi-oscillation 
measurement for different delay time ($\tau_{d}$) between the pump and qubit 
control.

Fig.~\ref{fig3}(c) shows the measurements of normalized integrated signal $V_H$
for different delays. The horizontal axis corresponds to the 
duration of the qubit control pulse. For comparison, a measurement made
in the absence of the pump pulse is included as well.
For $\tau_d=$~2.64~$\mu$s, we see small oscillations in the measurement 
indicating the coherent pouplation transfer between the ground and the first excited 
state. 
For such short delay time $\tau_d$, there are two effects that reduce the 
contrast of oscillations. First, the qubit population in the ground 
or first excited state could be low due to its excitation to higher
levels. Second, for short times, the pump photon occupancy in the
cavity can be substantial leading to dephasing.
We use slowly varying control pulse and therefore rule out any 
leakage of qubit to the higher levels by the control pulse. 
For $\tau_d=$~8.8~$\mu$s, the oscillations closely resemble the result 
obtained with pump maintained in off-condition.
We systematically measure the amplitude of Rabi oscillations 
for different delay time between pump and the control pulse. Fig~\ref{fig3}(d) 
shows the plot of oscillation amplitude for different delay times, showing the 
clear resurgence of the coherence in the device.
Additional dataset is included in the SI.
By fitting it to $\mathcal{F} (1 - e^\frac{t-t_0}{\tau})$, we extract the characteristic timescale, 
$t_0+\tau \approx$~4.8~$\mu$s for the relaxation from unconfined states.
We point out here that such a timescale involves contributions from the
relaxation of qubit from the higher excited states and 
from the dephasing due to occupancy of the cavity by pump-photons.

\begin{table*}[htb]
	\begin{tabular}{|p{6cm}|c|c|}
		\hline
		\textbf{Device parameter}    & \textbf{Symbol}   	& \textbf{Value}  \\ \hline
		Bare cavity frequency  & $\omega_c^0/2\pi$  & 6.002 GHz   \\ \hline
		Maximum qubit frequency  & $\omega_q^0/2\pi$  & 7.203 GHz  \\ \hline
		Kerr-nonlinearity  & $\alpha_q/2\pi$  & $-225$~MHz  \\ \hline
		Maximum Joesphson energy & $E_{J}^{0}/h$   & 30.65 GHz \\ \hline
		Cavity linewidth at zero flux  & $\kappa/2\pi$ & 1.38 MHz \\ \hline
		Qubit relaxation time at zero flux & $T_1$ & 2.11 $\mu$s\\ \hline
		Qubit cavity coupling & $g/2\pi$ & 87~MHz \\ \hline
	\end{tabular}
	\caption{Summary of device parameters studied in the main text}
	\label{table}
\end{table*}

After characterizing the response of the device under high power, next,
we perform the characterization of the flux bias port.
We utilize the high bandwidth of the flux drive line and create
a single-photon state in the cavity. 
The pulse protocol for such scheme is shown in Fig.~\ref{fig4}(a). It consists of initializing 
the qubit to the first excited state by applying a $\pi$-pulse. The qubit
frequency is then rapidly tuned to bring it in resonance with the cavity.
The modes are maintained in resonance for a variable time $\tau_{int}$ and then
the qubit mode is brought back to the original frequency followed by a measurement
pulse. During the interaction period, the qubit and the cavity modes exchange the 
single excitation coherently.
To understand the applicability of this scheme in a strongly driven system, 
we follow this protocol after a strong  pump (at the bare cavity 
frequency) pulse with varied delay time $\tau_d$.

The current in the flux loop, controlling the qubit detuning, is applied using an
arbitrary waveform generator. 
The shape of the flux-pulse is rectangular with rising  
and falling segments set as the half-gaussian with standard deviation of 1.1~ns. 
The width and the amplitude of the flux pulse are varied to control the 
interaction time and the qubit frequency, respectively.
Fig.~\ref{fig4}(b) shows $V_H$ as the interaction length is varied. 
For delay time of 8.8 $\mu$s ($>$ $t_0+\tau$), we observe that the qubit 
regains the coherence and oscillation due to the swapping of a single excitation
can be clearly seen.
Due to the strong coupling between the qubit and the cavity, it takes 
approximately 3~ns to transfer the single-photon from the qubit 
mode to the cavity mode indicated by the black dotted line.

Fig.~\ref{fig4}(c) shows the colorplot of $V_H$ as the qubit 
detuning and interaction duration is varied. The oscillation frequency 
of the single excitation swap changes as the relative detuning between the qubit 
and cavity mode frequency is varied.
At zero detuning, the oscillation frequency is minimum at 2$g$ and increases 
to $\sqrt{4 g^2 + \Delta^2}$ with detuning \cite{oconnell_quantum_2010}. 
The deviation from the ideal chevron pattern suggests that the flux-pulse
disperses as it travels down the sample. The initial change in the flux pulse 
is not able to tune the qubit in resonance with the cavity for short interaction 
time. We observe a small distortion in the chevron pattern for time-scales 
shorter than 10~ns. From this deviation, we concluded the bandwidth of the flux-line to 
be approximately 100~MHz. The bandwidth of flux line is limited by the 
parasitic capacitance and self-inductance of the current loop patterned 
near the SQUID loop. While we try to maintain a 50 $\Omega$ environment 
till the connector on the cavity, the impedance of the flux line on the 
silicon chip deviates from 50 $\Omega$ and this limits the bandwidth.
Such distortions in flux-pulse, in principle, could be improved by using 
pre-compensated flux-pulses. To better understand the experimental 
results, we numerically simulate the system 
by solving the Lindblad master equation with the flux pulse sequence used in the 
experiment \cite{johansson_qutip_2012}.
The simulated outcome of the excited state population is plotted 
in Fig.~\ref{fig4}(d) with 
variable detuning in the vertical axis. 
The difference between the simulation and experimental 
plots can be understood from the distorted flux-pulse at the sample, 
as discussed above.

To summarize, we demonstrated a design of a fast-tunable transmon qubit in 
a 3D waveguide cavity architecture. We characterized its relaxation from unconfined 
states to the ground state after a high power drive pulse.
We measure a resurgence time of 4.8~$\mu$s. We characterize the fast-flux line
and find a bandwidth of $\approx$~100~MHz.
These performance benchmarking results provide the design 
guidelines for hybrid systems intended to integrate additional degrees of freedom 
with the circuit-QED platform \cite{gunupudi_optomechanical_2019, peterson_ultrastrong_2019}.

\section*{Acknowledgment}

This material is based upon work supported by the Air Force Office of 
Scientific Research under award number FA2386-20-1-4003. V.S. acknowledge the 
support received under the Young Scientist Research Award by the Department of 
Atomic Energy and support received under the Core Research Grant by the Department of 
Science and Technology (India). The authors acknowledge device fabrication 
facilities at CeNSE, IISc
Bangalore, and central facilities at the Department of
Physics funded by DST.

\end{document}